\documentclass[traditabstract]{aa}
\usepackage{graphicx}
\usepackage{txfonts}
\usepackage{natbib}
\bibpunct{(}{)}{;}{a}{}{,}

\newcommand{\Msun}{\ensuremath{\,{\rm M}_\odot}}           
\newcommand{\logg}{\ensuremath{\log g}}                    
\newcommand{\kms}{\,km\,s$^{-1}$}                          
\newcommand{\Apx}{\,\AA\,px$^{-1}$}                        
\newcommand{\mc}[1]{\multicolumn{2}{c}{#1}}
\newcommand{\as}{\ensuremath{^{\prime\prime}}}             

\begin{document} 

\title{Orbital periods of cataclysmic variables identified by the SDSS.}

\subtitle{VIII. A slingshot prominence in SDSS J003941.06$+$005427.5?}

\author{John Southworth \and T.\ R.\ Marsh \and B.\ T.\ G\"ansicke \and D.\ Steeghs \and C.\ M.\ Copperwheat}

\institute{Department of Physics, University of Warwick, Coventry, CV4 7AL, UK \ \ \ \ \ \email{jkt@astro.keele.ac.uk}}

\date{Received 24 August 2010; accepted 27 September 2010}       

\abstract{We present VLT spectroscopy and NTT photometry of the faint cataclysmic binary SDSS J003941.06$+$005427.5. This object shows triple-peaked H$\alpha$ emission with all three peaks variable in both strength and velocity. We measure an orbital period of $91.395 \pm 0.093$\,min from the velocity variations of the wings of the H$\alpha$ emission line. Using the GALEX and SDSS photometry of this object, we determine a white dwarf temperature of $15000$\,K and a very late ($\ga$L2) spectral type for the companion star. These measurements, plus the relatively long orbital period, suggest that SDSS J003941.06$+$005427.5 may be a post-bounce cataclysmic variable. Doppler maps of the H$\alpha$ and He\,I 6678\,\AA\ emission features show an accretion disc with a non-uniform brightness and departures from Keplerian flow. The third emission peak is detected only in H$\alpha$ and at a relatively low velocity amplitude of $202 \pm 3$\kms. We are unable to explain this emission as arising from either the white dwarf, the secondary star, or the accretion disc. We tentatively attribute this mysterious central peak to a coronal loop anchored at the secondary star. If confirmed, this would be the first example of a slingshot prominence in a CV with a low mass-transfer rate and/or a fully convective secondary star.}


\maketitle 

\section{Introduction}                                                                       \label{sec:intro}

Cataclysmic variables (CVs) are interacting binaries composed of a low-mass secondary star which is transferring material to a white dwarf (WD) primary star. In most cases the mass donor is unevolved and the mass transfer occurs via an accretion disc surrounding the WD. Detailed reviews of the properties of CVs have been given by \citet{Warner95book} and \citet{Hellier01book}.

We are undertaking a project to characterise the population of CVs identified by the Sloan Digital Sky Survey (SDSS) \citep{Gansicke+09mn, Me+07mn, Me+07mn2, Me+09aa, Me+10aa, Dillon+08mn}. In the course of this work we obtained time-resolved spectroscopy and photometry of the faint system SDSS\,J003941.06$+$005427.5 (hereafter SDSS\,J0039) which shows strong H$\alpha$ emission with a unique variable triple-peaked structure. Doppler maps of the spectroscopic emission lines suggest that not only is the accretion disc markedly non-circular, but that there is an inner emission feature which defies explanation in the standard framework of the structure of CVs.

SDSS\,J0039 was discovered to be a CV by \citet{Szkody+05aj} on the basis of an SDSS spectrum which shows Balmer and He\,I emission lines. Very broad absorption is visible around the higher-order Balmer lines, indicating that the WD emits a large fraction of the overall light of the system.



\section{Observations and data reduction}                                           \label{sec:obs}

\begin{table*} \centering
\caption{\label{tab:obslog} Log of the observations presented in this
work. The acquisition magnitudes were measured from the VLT/FORS2
acquisition images and are discussed in Sect.\,\ref{sec:obs:vltspec}.}
\begin{tabular}{lccccccc} \hline
Date&Start time&End time&Telescope and&Optical& Number of  &Exposure&Mean     \\
(UT)&  (UT)    &  (UT)  &  instrument &element&observations&time (s)&magnitude\\
\hline
2007 08 07 & 09:02 & 10:38 & NTT\,/\,SUSI2 & unfiltered  & 111 &  60 & 20.7 \\
2007 08 08 & 09:25 & 10:32 & NTT\,/\,SUSI2 & unfiltered  &  55 &  60 & 20.6 \\[2pt]
2007 08 15 & 06:20 &       & VLT\,/\,FORS2 & unfiltered  &   1 &  10 & 20.8 \\
2007 08 15 & 06:24 & 10:20 & VLT\,/\,FORS2 & 1200R grism &  22 & 600 & --   \\
2007 08 16 & 04:15 &       & VLT\,/\,FORS2 & $V$ filter  &   1 &  30 & 20.7 \\
2007 08 16 & 04:20 & 05:32 & VLT\,/\,FORS2 & 1200R grism &   7 & 600 & --   \\
\hline \end{tabular} \end{table*}

\subsection{VLT spectroscopy}                                                       \label{sec:obs:vltspec}

Spectroscopic observations were carried out in 2007 August using the FORS2 spectrograph \citep{Appenzeller+98msngr} at the Very Large Telescope (VLT), Chile (Table\,\ref{tab:obslog}). The 1200R grism was used, giving a wavelength coverage of 5870\,\AA\ to 7370\,\AA\ with a reciprocal dispersion of 0.73\Apx\ and a resolution of 1.6\,\AA\ at H$\alpha$ with a slit width of 1\as\ and the CCD binned 2$\times$2.

The data were reduced using optimal extraction \citep{Horne86pasp} as implemented in the {\sc pamela}\footnote{{\sc pamela} and {\sc molly} were written by TRM and can be found at {\tt http://www.warwick.ac.uk/go/trmarsh}}
code \citep{Marsh89pasp}, which also makes use of the {\sc starlink}\footnote{The Starlink Software Group homepage can be found at {\tt http://starlink.jach.hawaii.edu/}} packages {\sc figaro} and {\sc kappa}.
The wavelength calibration of the spectra was performed using one arc lamp exposure for each night. Wavelength shifts due to spectrograph flexure were measured and accurately removed using the 6300.304\,\AA\ night sky emission line \citep[see][]{Me+06mn,Me+08mn,Me++08mn}.

The observing procedure of FORS2 included obtaining target acquisition images from which photometry can be obtained. The first of these images was taken unfiltered whereas the second was obtained with a $V$ filter. We extracted differential photometry from these images using the {\sc starlink} package {\sc gaia}. The $V$-band apparent magnitudes of the comparison stars were calculated from their $g$ and $r$ magnitudes using the transformations provided by \citet{Jester+05aj}.

\subsection{NTT photometry}                                                         \label{sec:obs:ntt}

Time-series photometry was obtained during 2007 August using the New Technology Telescope (NTT) at ESO La Silla, Chile, and the SUSI2 CCD mosaic imager \citep{Dodorico+98spie}. CCD\#45 was used, binned by a factor of three in both directions to give a plate scale of 0.24\as\,px$^{-1}$. The observations were performed in white light in order to maximise throughput, and care was taken to place the target and comparison stars on parts of the CCD which were least affected by fringing. Exposure times of 60\,s gave an observing cadence of 75\,s.

Debiasing and flat-fielding of the raw images was performed with the {\sc starlink} software packages {\sc convert} and {\sc kappa}. Optimal and aperture photometry \citep{Naylor98mn} was obtained from the reduced images with the {\sc multiphotom} script \citep{Me++04mn}, which uses the {\sc starlink autophotom} package \citep{Eaton++99}. The differential photometry was shifted to an apparent magnitude scale using the mean of the $g$ and $r$ magnitudes of the comparison star.

\section{Results}                                                                   \label{sec:results}

\subsection{Orbital period measurement}                                             \label{sec:porb}

\begin{table} \centering
\caption{\label{tab:rvorbit} Best-fitting spectroscopic orbit found
using {\sc sbop}. Phase zero corresponds to inferior conjunction.}
\begin{tabular}{l r@{\,$\pm$\,}l}\hline
Reference time (HJD)          & 2454327.78211 & 0.00038     \\
Orbital period (d)            &     0.0634686 & 0.000064    \\
Eccentricity                  &      \mc{0.0 (fixed)}       \\
Velocity amplitude (\kms)     &         139.2 & 4.2         \\
Systemic velocity (\kms)      &        $-$8.2 & 3.1         \\
$\sigma_{\rm rms}$ (\kms)     &         \mc{16.3}           \\
\hline \end{tabular} \end{table}

\begin{figure} \centering
\includegraphics[width=0.48\textwidth,angle=0]{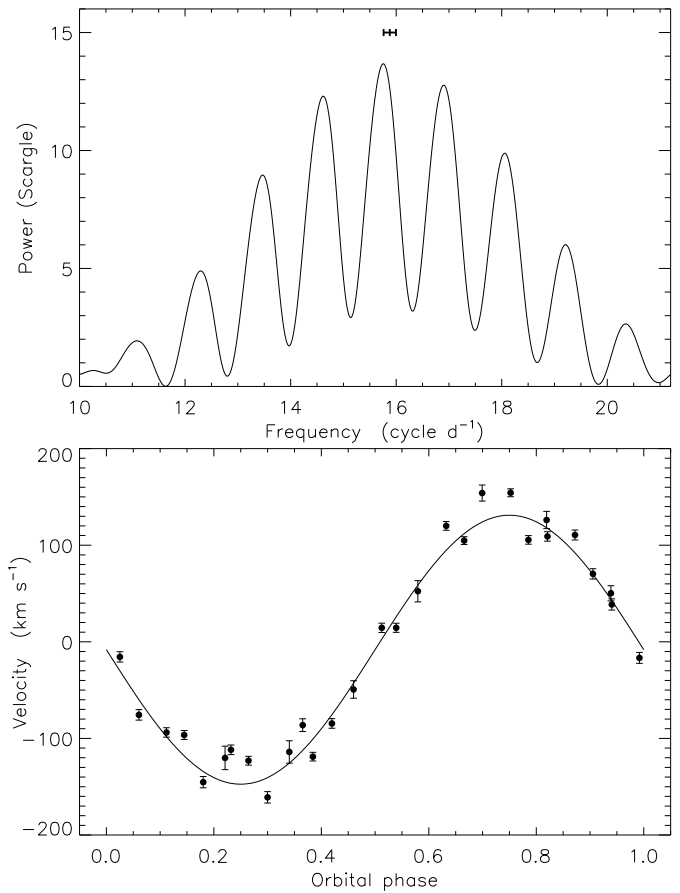} \\
\caption{\label{fig:0039:rvplot} {\it Upper panel:} Scargle periodogram of
the radial velocities of SDSS\,J0039 measured using a double Gaussian with
widths 300\kms\ and separation 2000\kms. The measured period and uncertainty
from the first night's observations alone are indicated with a thick line.
{\it Lower panel:} measured radial velocities (filled circles) compared to
the best-fitting spectroscopic orbit (unbroken line). The errorbars include
only the Poisson noise contribution to the velocity measurements.} \end{figure}

\begin{figure} \centering \includegraphics[width=0.48\textwidth,angle=0]{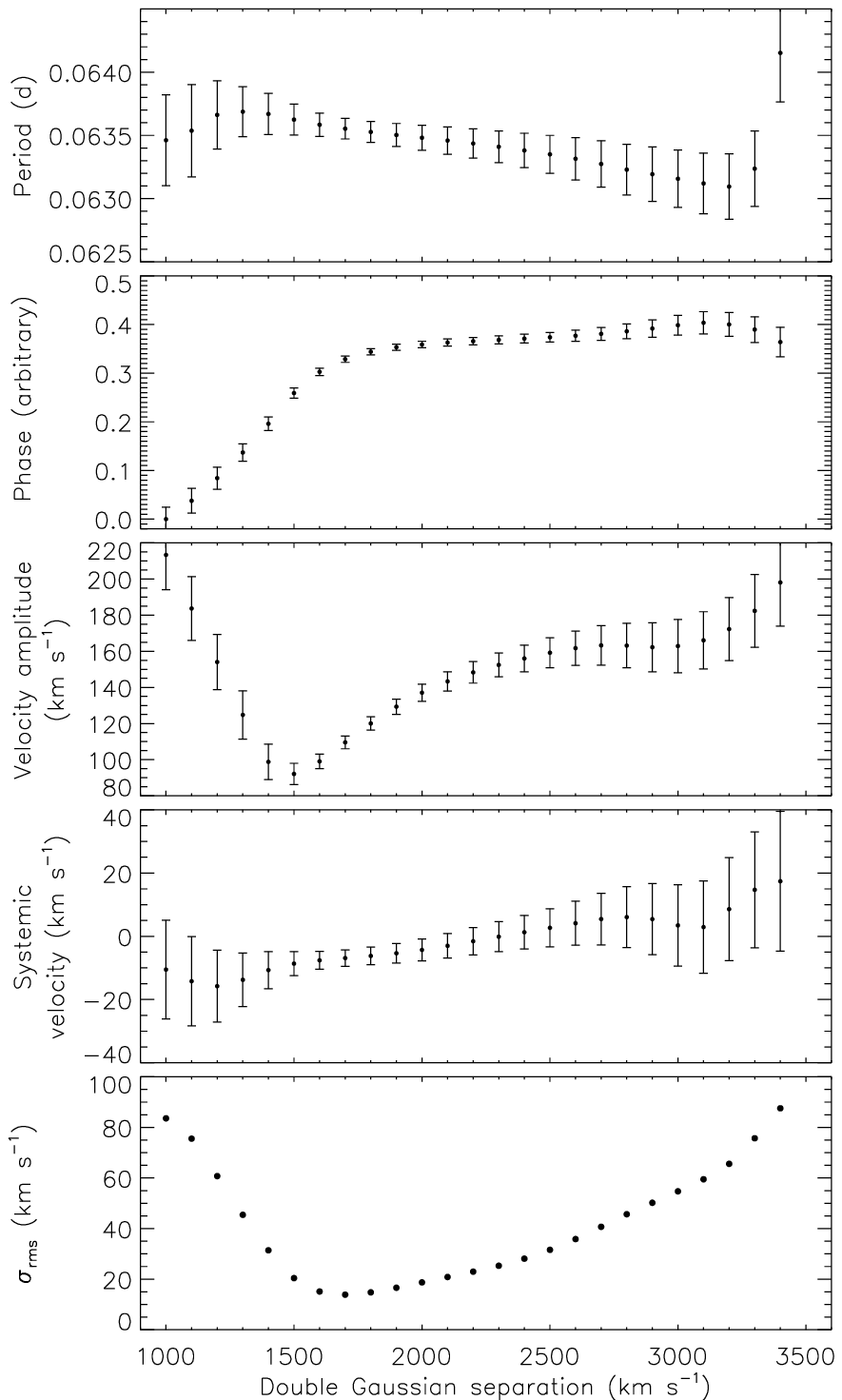} \\
\caption{\label{fig:diagnostic} Diagnostic diagram showing the variation of the
best-fitting spectroscopic orbital parameters for radial velocities measured with a
range of separations using the double Gaussian function. In the absence of physical
constraints, the orbital phase has arbitrarily been defined to be zero for the lowest
Gaussian separation. $\sigma_{\rm rms}$ denotes the scatter of the velocity measurements
around the fitted orbit.} \end{figure}

We obtained 22 VLT spectra of SDSS\,J0039 on the night of 2007 August 15th and a further 7 spectra on the next night. We have measured the profiles for radial velocity motion by cross-correlation against single and double Gaussian functions \citep{SchneiderYoung80apj,Shafter83apj}, as implemented in {\sc molly}. The best results were obtained using a double Gaussian with widths 300\kms\ and separation $\xi = 2000$\kms. We have fitted a spectroscopic orbit to the radial velocities using the {\sc sbop}\footnote{Spectroscopic Binary Orbit Program, written by P.\ B.\ Etzel, \\ {\tt http://mintaka.sdsu.edu/faculty/etzel/}} code, which gives reliable error estimates for the optimised parameters \citep{Me+05mn}.

The radial velocities from the first night of VLT observations give an unambiguous period measurement of $90.69 \pm 0.67$\,min, and including the seven spectra from the second night gives a refined period of $91.395 \pm 0.093$\,min. The one-day alias solutions at 85.2 and 98.5 min can be ruled out, as they strongly disagree with the period from the data taken only on the first night and lead to radial velocity curves with a much increased scatter. The parameters of the best-fitting orbit are given in Table\,\ref{tab:rvorbit}. A periodogram \citep{Scargle82apj} and phased radial velocity curve for the full dataset are shown in Fig.\,\ref{fig:0039:rvplot}.

The velocity amplitude of this spectroscopic orbit, $K_{\rm em} = 139.2 \pm 4.2$\kms\ has been measured from the wings of the H$\alpha$ emission line and is unlikely to accurately represent the motion of the WD: accretion disc asymmetries can induce spuriously large RV excursions even in the wings of emission lines. To investigate the reliability of the measurement we have constructed a diagnostic diagram (see \citealt{Shafter++86apj} and \citealt{Thorstensen00pasp}), where orbital solutions are plotted for a range of $\xi$ values (Fig.\,\ref{fig:diagnostic}). As SDSS\,J0039 is not eclipsing, the time of zero orbital phase is arbitrary. The results for small $\xi$ values are very unreliable, but solutions with $1800 < \xi < 3000$\kms\ are in mutual agreement and should best represent the motion of the inner accretion disc. For these $\xi$ values we find $120 < K_{\rm em} < 165$\kms.

\begin{figure*} \centering
\includegraphics[width=0.32\textwidth,angle=0]{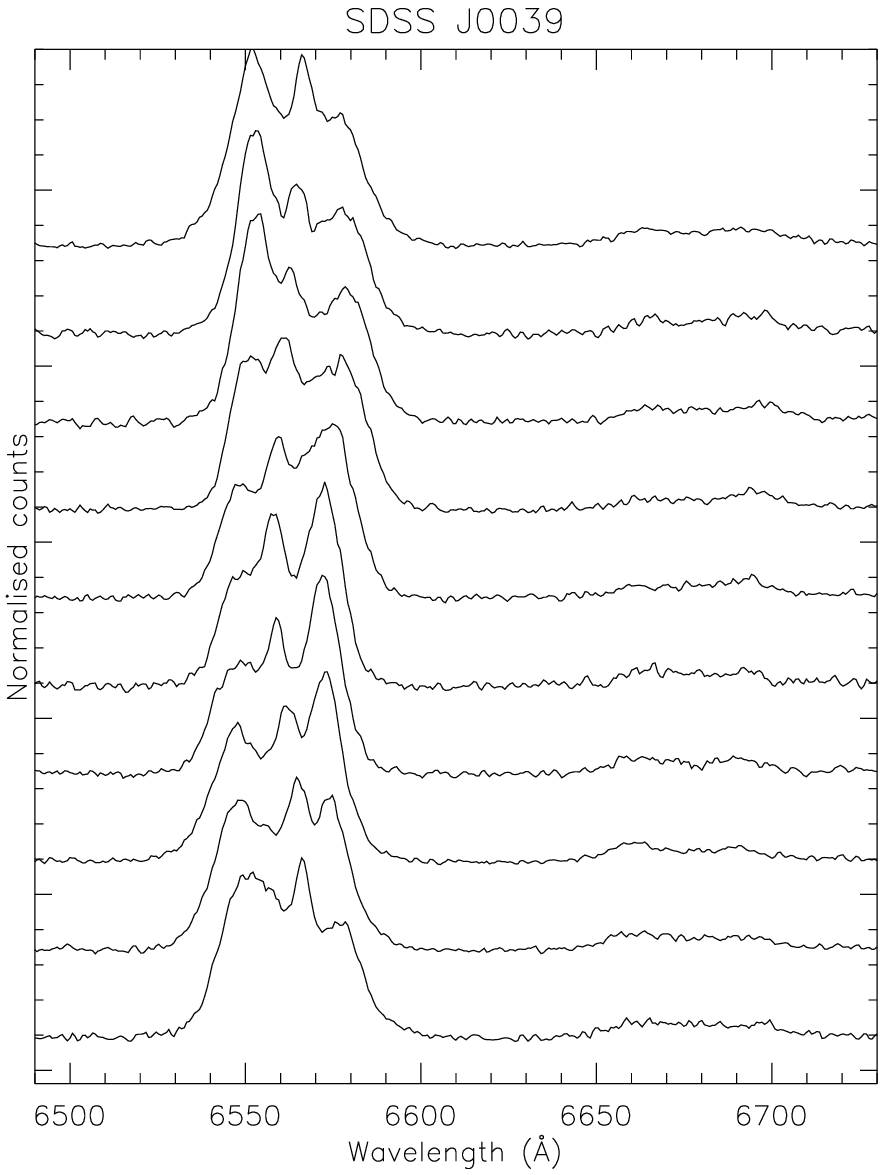}
\includegraphics[width=0.32\textwidth,angle=0]{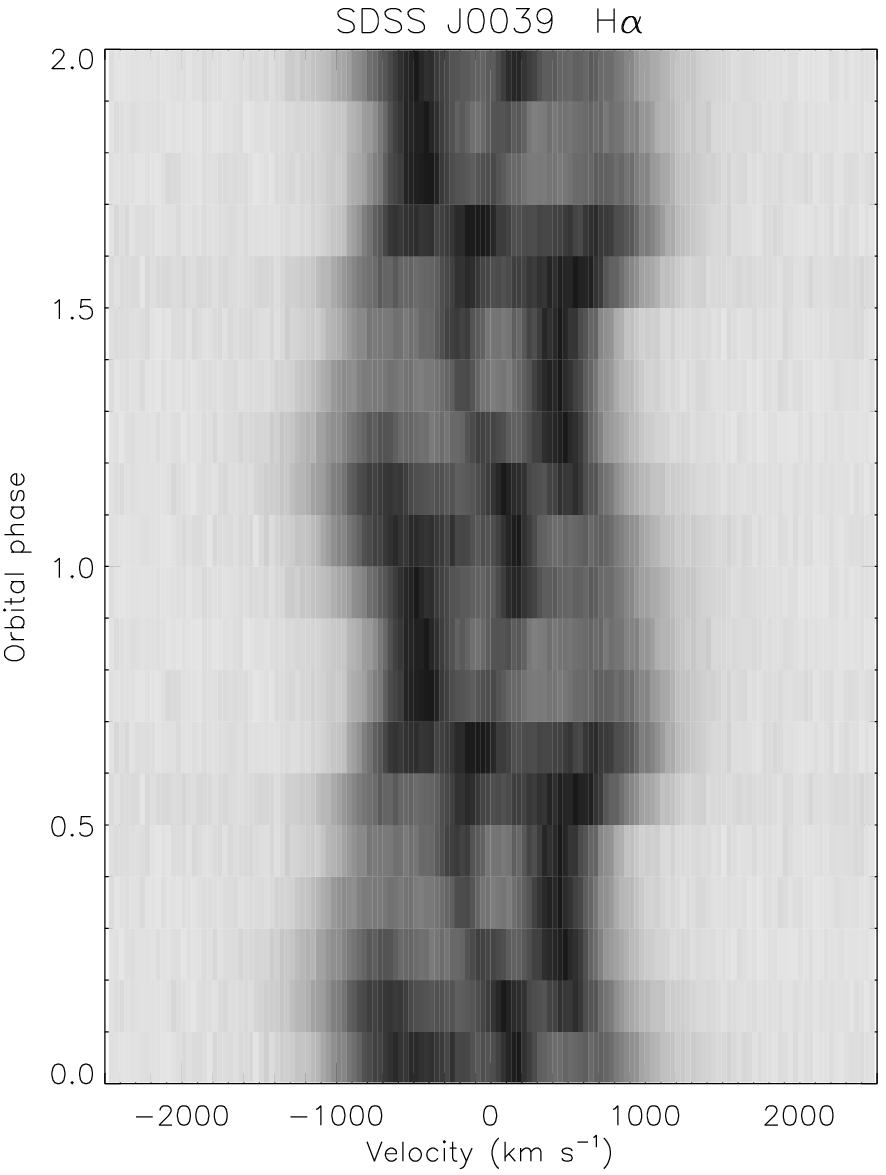}
\includegraphics[width=0.32\textwidth,angle=0]{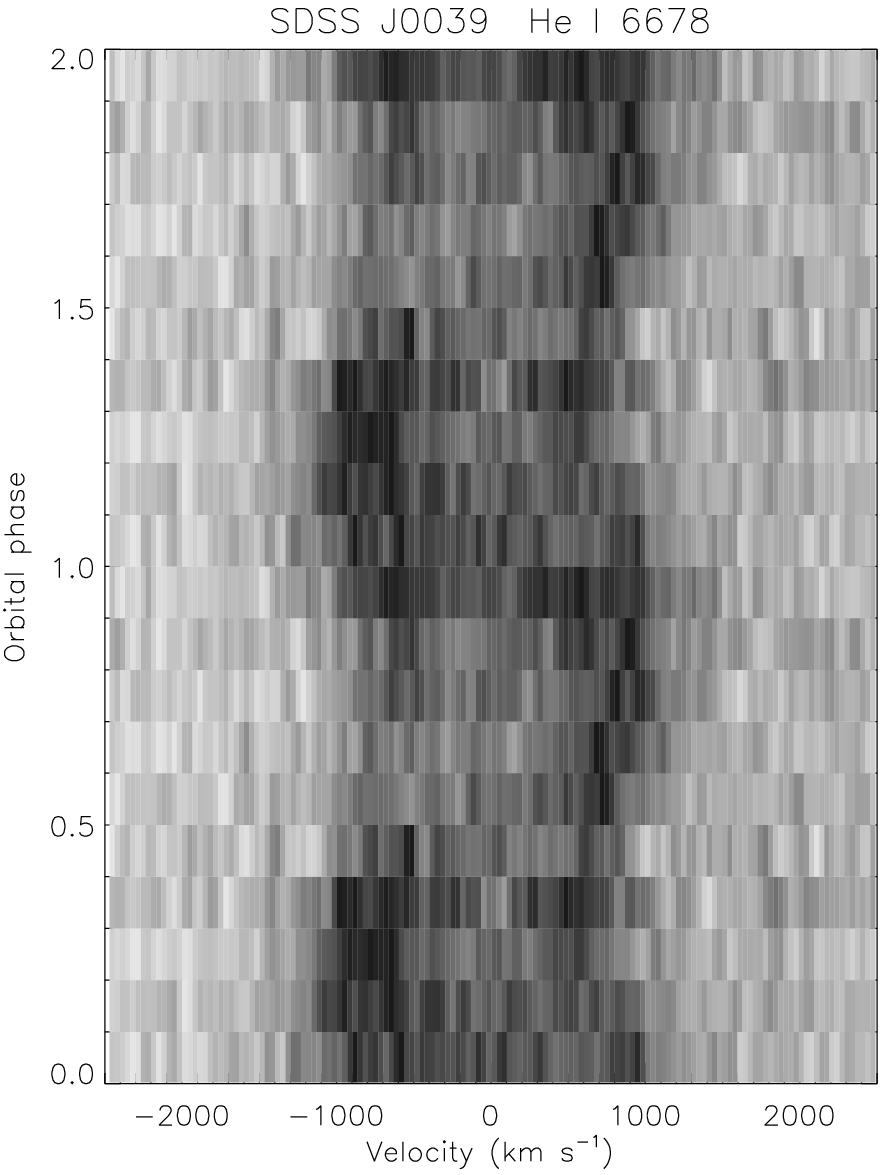}
\caption{ \label{fig:0039:stacked} \label{fig:spec:trailed}
{\it (1)} Phase-binned and stacked spectra of SDSS\,J0039. The spectra
have been continuum-normalised and plotted with offsets of +3.
{\it (2)} Phase-binned and trailed spectra in the region of H$\alpha$.
{\it (3)} Phase-binned and trailed spectra in the region of the He\,I
6678\,\AA\ emission line. The He\,I spectra have been smoothed with a
Savitsky-Golay filter for display purposes.} \end{figure*}

Using an orbital period of 91.395\,min, we have combined the spectra into ten phase bins. Fig.\,\ref{fig:0039:stacked} shows the binned spectra stacked (for H$\alpha$) and trailed (H$\alpha$ and He\,I 6678\,\AA). The structure of the H$\alpha$ profile of SDSS\,J0039 immediately stands out as unusual in that it contains {\em three} separate emission features: double peaks characteristic of a high-inclination accretion disc, and a central pinnacle which is always present and clearly differentiable from the other two peaks.

On closer inspection several others oddities are discernable. Firstly, the variable strength of the outer double peaks is not the expected behaviour of an accretion disc in quiescent CVs. Secondly, the velocity variation of the central peak is not in phase with this variation in strength of the double peaks. Thirdly, the He\,I 6678\,\AA\ and 7065\,\AA\ lines do not exhibit the central emission feature. The helium lines {\em do} have double peaks with variable strength, but this variability is not in phase with the modulation seen in H$\alpha$.

\subsection{Photometric properties}                                                 \label{sec:phot}

\begin{figure} \centering \includegraphics[width=0.48\textwidth,angle=0]{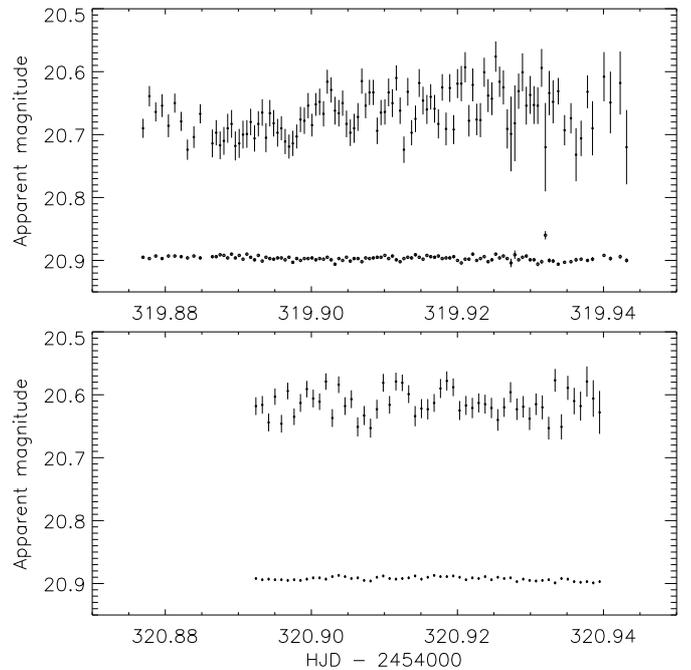} \\
\caption{\label{fig:0039:lcplot} NTT unfiltered photometry of SDSS\,J0039 obtained
over two nights (shown in separate panels). Differential magnitudes for SDSS\,J0039
minus comparison star are shown as filled circles with error bars, offset by the $V$
magnitude of the comparison. Differential magnitudes for comparison minus check star
are shown offset to appear at the bottom of the plot; in most cases the errorbars are
smaller than the points.} \end{figure}

2.7 hours of unfiltered photometry of SDSS\,J0039 was obtained over two nights using NTT/SUSI2 (fig.\,\ref{fig:0039:lcplot}). We have calculated \citet{Scargle82apj}, AoV \citep{Schwarzenberg89mn} and ORT \citep{Schwarzenberg96apj} periodograms of the NTT light curves, and have found no evidence of eclipses or any other periodic variability. The light curves display a scatter which is greater than the measurement errors. This indicates that flickering is present, as is common in CVs \citep[e.g.][]{Bruch92aa,Bruch00aa}.

The mean magnitudes of SDSS\,J0039 are 20.7 and 20.6 on the two night of NTT observations, which is very similar to the magnitudes in the VLT/FORS2 acquisition images (20.8 and $V = 20.7$; Sect.\,\ref{sec:obs:vltspec}) and to the original SDSS photometric and spectroscopic observations ($g = 20.56$ and $r_{\rm spec} = 20.67$). In addition to this, SDSS\,J0039 was observed at 29 epochs over the years 1999--2003 in the course of the SDSS variability study of `Stripe 82' \citep{Bramich+08mn} and was always found to be close to $g = 20.6$. The low photometric variability and large emission line strengths indicate that SDSS\,J0039 was in quiescence during every observation detailed above, and outbursts only rarely.

\subsection{Spectral energy distribution}                                           \label{sec:sed}

\begin{figure} \centering \includegraphics[angle=-90,width=0.48\textwidth]{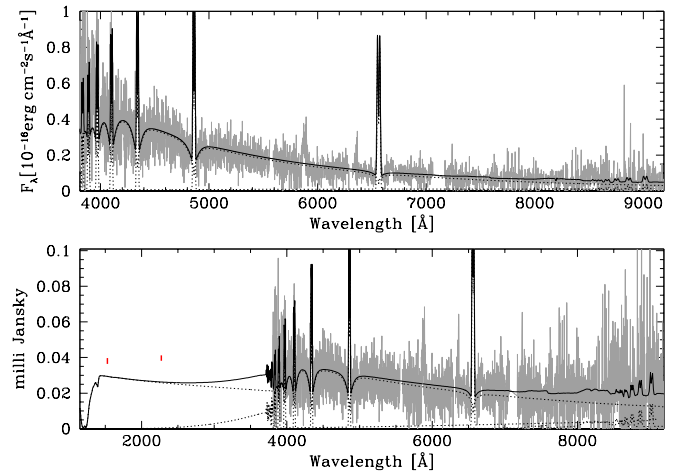} \\
\caption{\label{fig:0039:sed} Three-component model of the SDSS spectrum of SDSS\,J0039. The
three components are a WD with $T_{\rm WD} = 15\,000$\,K, $R_{\rm WD} = 8.7 \times 10^6$\,m;
an isothermal and isobaric hydrogen slab, and an L2 secondary star, all scaled to a distance
of $d = 790$\,pc. The two red points represent data from the GALEX satellite.} \end{figure}

We have studied the SDSS spectrum and GALEX ultraviolet fluxes of SDSS\,J0039 \citep{Morrissey+07apjs} using the model described in \citet{Gansicke+99aa}, \citet{Rodriguez+05aa} and \citet{Gansicke+06mn}, which includes flux contributions from the WD, the secondary star, and an isothermal and isobaric hydrogen disc. The surface gravity of the WD was fixed at $\logg = 8$ (cgs). We find a best fit for a WD effective temperature of $T_{\rm WD} =15\,000$\,K and a distance of 790\,pc (Fig.\,\ref{fig:0039:sed}). The spectral type of the mass donor is constrained to be later than approximately L2. This best-fitting model under-predicts the GALEX fluxes, but increasing $T_{\rm WD}$ would compromise the fit to the SDSS spectrum. There may be some additional component in the UV making up 20\% of the flux, but this suggestion is entirely dependent on the absolute flux calibrations of the SDSS spectrum and the GALEX fluxes.

\subsection{SDSS\,J0039: a post-bounce CV candidate?}

In their study of the SDSS CV population as a whole, \citet{Gansicke+09mn} showed that a characterstic feature of this sample is an accumulation of intrinsically faint CVs in the period range 80--86\,min. The vast majority of these systems have optical spectra dominated by the emission from their white dwarfs, with no noticeable contribution from the secondary stars, implying extremely late secondary spectral types. These findings are consistent with theoretical predictions that the companion stars are whittled down as the CVs evolve towards shorter orbital periods, reducing the mass below the hydrogen burning limit at the minimum period, after which evolution proceeds slowly back to longer periods \citep{Kolb93aa,KolbBaraffe99mn}.

Whilst a number of CV secondary stars with sub-stellar masses have been identified near the minimum period \citep{Littlefair+06sci,Littlefair+08mn}, there is yet no unimpeachable detection of the predicted ``post-bounce'' population. The expected characteristics of post-bounce CVs are a very low accretion rate, a low WD temperature, a very low secondary star mass, and a brown-dwarf secondary spectral type. The ability to differentiate between pre- and post-bounce CVs increases with longer orbital periods, because of the greater difference in secondary spectral type and mass transfer rate between the two types of systems.

Among the SDSS CVs that have white-dwarf dominated optical spectra, SDSS\,J0039 exhibits currently the best credentials to be a post-bounce system, as it has an extremely late-type companion star ($\ga$L2) and an orbital period clearly beyond the 80--86\,min period spike. The post-bounce nature of our object can be tested by near-infrared photometry, which would provide a stronger constraint on the spectral type of the companion star.

\subsection{Doppler tomography}

\begin{figure*} \centering \includegraphics[height=\textwidth,angle=270]{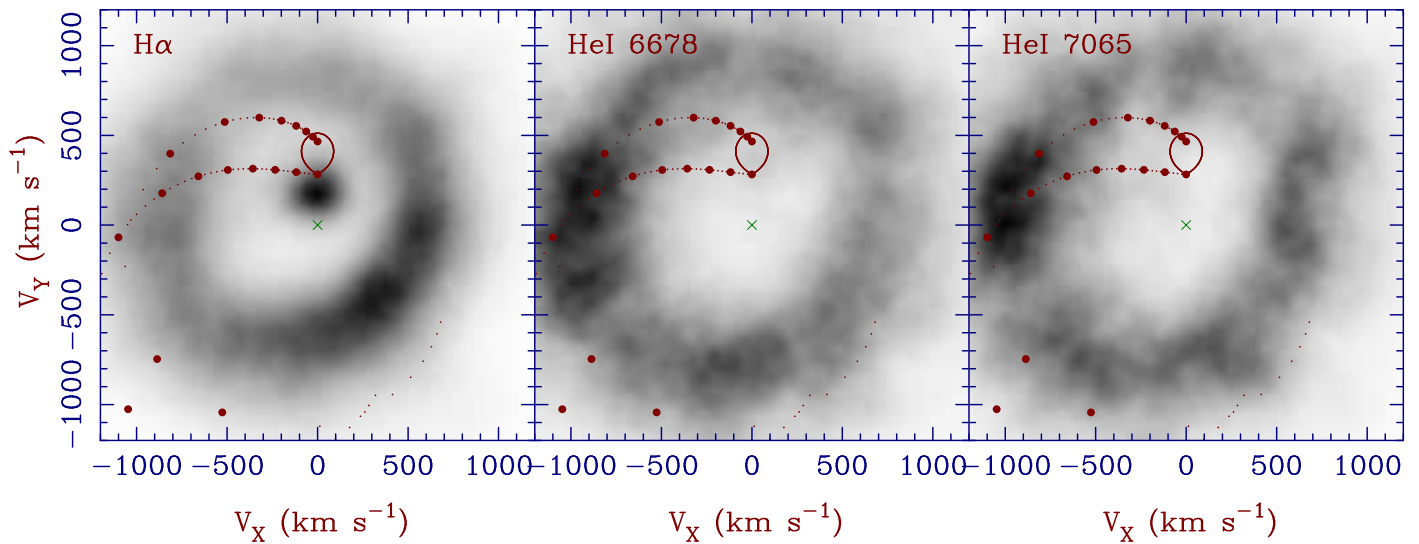} \\
\caption{\label{fig:0039:map} Doppler maps of the SDSS\,J0039 system for the H$\alpha$
(left), He\,I 6678\,\AA\ (centre) and He\,I 7065\,\AA\ (right) emission lines. The
Roche lobe of the mass donor is indicated by a solid curve and the centre of mass of
the system by a cross. The upper dots indicate the velocity of the accretion stream
and the lower dots the Keplerian velocity of the disc along the path of the stream.
They are positioned at every $0.1R_{\rm RL}$, decreasing from $R_{\rm RL} = 1.0$ at
the mass donor, where $R_{\rm RL}$ is the radius of the WD Roche lobe. This
configuration corresponds to $K_{\rm WD} = 60$\kms\ and $K_2 = 400$\kms.} \end{figure*}


Doppler maps of the H$\alpha$, He\,I 6678\,\AA\ and He\,I 7065\,\AA\ emission lines were computed using the maximum entropy method \citep{MarshHorne88mn}. We have overlaid a basic interpretation of SDSS\,J0039 onto the Doppler maps (Fig.\,\ref{fig:0039:map}). The centre of mass of the system is shown with a cross, and the expected position of the surface of the secondary star in velocity space is shown with an unbroken line. The dots and dotted lines indicate the velocity of the accretion stream and the Keplerian velocity of the disc along the path of the stream.

The emission characteristics apparent in the trailed spectra of SDSS\,J0039 translate into unusual features in the Doppler maps. The first of these is the intense inner emission feature, which is clearly visible at H$\alpha$ but totally absent in the He\,I lines. The maps have been rotated to bring this feature onto the line passing through the centres of mass of the two stars, the standard configuration for Doppler maps. In the next section we consider and reject the possibility that the inner emission feature arises from the surface of either the WD or secondary star.

The Doppler maps bring out another extraordinary feature of SDSS\,J0039 which is less obvious in the data than the central S-wave. The outer disc in H$\alpha$ appears to be extremely non-circular, given that circular motion leads to symmetry around the centre of mass of the WD, located just below the centre of mass of the system. The non-circularity is extreme: the bright region in the lower-right quadrant has a velocity a factor of two lower than that of the upper-left quadrant. A Keplerian velocity profile ($V \propto r^{-1/2}$) then implies that the outer disc varies in radius by a factor of four! The behaviour of the He\,I lines is very different and more symmetric, which suggests that the lower-right quadrant of H$\alpha$ is in some way unusual. We can only interpret this feature in the Doppler maps as evidence of non-Keplerian flow.

\section{The origin of the central spike}

\begin{figure} \centering \includegraphics[height=0.48\textwidth,angle=270]{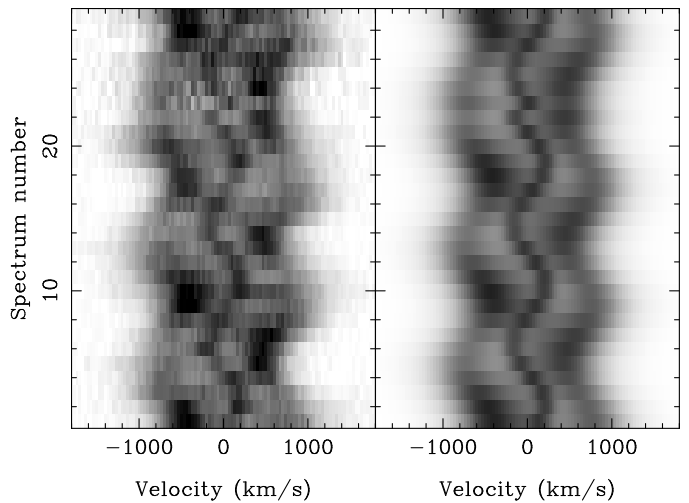} \\
\caption{\label{fig:0039:fitplot} Trailed representation of the 29 observed spectra (left)
compared to the best-fitting representation for the four-Gaussian fit (right).} \end{figure}

The inner emission feature is unusual in normal CVs and reminiscent of the ``central spike'' seen in some AM\,CVn (ultracompact binary) stars, where it is ascribed to the WD (e.g.\ GP\,Com; \citealt{Morales+03aa}). To investigate this we developed a six-Gaussian model to fit the observed H$\alpha$ line profiles. The two wide emission peaks from the accretion disc are each modelled with one wide and one narrow Gaussian, the central spike with a fifth Gaussian, and the bright spot with a sixth. Each Gaussian was allowed to vary sinusoidally in both radial velocity and brightness on the orbital period. The fit is plotted in Fig.\,\ref{fig:0039:fitplot}.

We find a velocity amplitude of $K_{\rm spike} = 202 \pm 3$\kms\ for the central spike, which immediately rules out the `AM\,CVn explanation' as it is too high to be associated with the WD. On the other hand it is much too low to be from the bright spot region on the accretion disc, which the Doppler maps show to have a velocity of about 800--900\kms. The only stable feature of the system left is the secondary star, so we now consider this possibility.

\subsection{The whole secondary star?}

\begin{figure} \includegraphics[width=0.48\textwidth,angle=0]{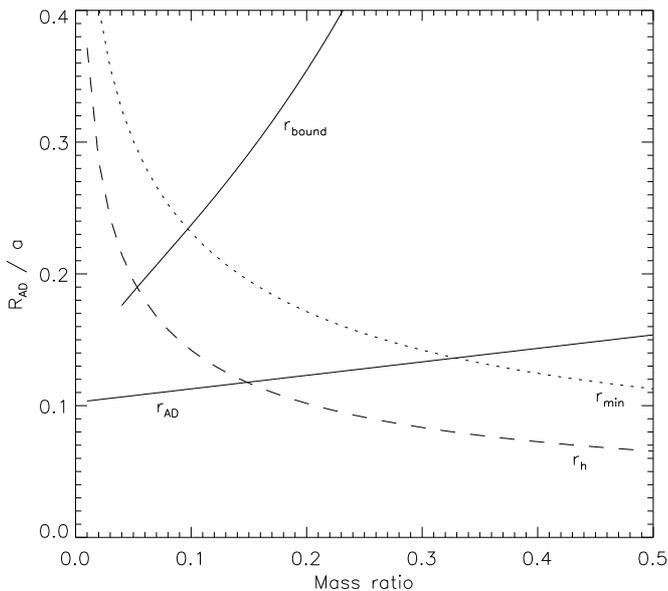} \\
\caption{\label{fig:qconstraints} Constraints on the mass ratio ($q$) obtained
from the size of the accretion disc. $r_{\rm AD}$ shows the accretion disc radius
according to Eq.\,\ref{eq:r_ad} and adopting $K_2 = 202$\kms. $r_{\rm bound}$ shows
how $r_{\rm AD}$ increases if the measured $K_2$ actually represents the kinematics
of the L1 point. $r_{\rm h}$ and $r_{\rm min}$ are the circularisation and minimum
radii of the gas stream.} \end{figure}

It is very unlikely that the central spike originates from the whole surface of the secondary, because $K_{\rm spike}$ is very low given that the velocity amplitude of the accretion disc emission peaks is $V_{\rm AD} = 631$\kms. Consider the following relations:
\begin{equation} K_2 = \frac{\Omega a M_{\rm WD} \sin i}{M_{\rm WD}+M_2} \end{equation}
\begin{equation} V_{\rm AD} = \sqrt{\frac{GM_{\rm WD}}{R_{\rm AD}}} \sin i \end{equation}
under the assumption of Keplerian velocities. The orbital angular frequency, $\Omega$, is given by
\begin{equation} \Omega^2 = \frac{G(M_{\rm WD}+M_2)}{a^3} \end{equation}
where $a$ and $i$ are the orbital separation and inclination, $M_{\rm WD}$ and $M_2$ are the masses of the component stars, $K_2$ is the velocity amplitude of the centre of mass of the secondary star, and $R_{\rm AD}$ is the radius of the outer accretion disc. From this we find
\begin{equation} \frac{R_{\rm AD}}{a} = r_{\rm AD} = (1+q) \left(\frac{K_2}{V_{\rm AD}}\right)^2 \label{eq:r_ad} \end{equation}
which for the values quoted above gives
\begin{equation} r_{\rm AD} = 0.10 (1+q) \end{equation}
where $q = {M_2}/{M_{\rm WD}}$ is the mass ratio and $a$ is the semimajor axis. This value can be compared with the minimum radius ($r_{\rm min} = R_{\rm min}/a$) reached by the gas stream if there is no disc (\citealt{Nelemans+01aa}, eq.\,6), and with the circularisation radius ($r_{\rm h} = R_{\rm h}/a$) which is the radius of the disc when its specific angular momentum matches that input from the stream (\citealt{VerbuntRappaport88apj}, eq.\,13 and allowing for their inverted definition of $q$). Adopting a reasonable $q = 0.15$ for the orbital period of SDSS\,J0039 \citep{Knigge06mn} yields $r_{\rm min} = 0.12$ and $r_{\rm h} = 0.19$. This is shown in Fig.\,\ref{fig:qconstraints}. Thus if the spike faithfully represents $K_2$, the radius implied for the disc, $r_{\rm AD} = 0.115$, is considerably less than the circularisation radius, and a little less even than the minimum radius that the gas stream can reach. The circularisation radius is the smallest radius that the accretion disc can have, and is usually comfortably less than the measured radius. Here it would have to be the other way around, so we reject the possibility that $K_{\rm spike} = K_2$.

\subsection{The inner Lagrangian point?}

If the emission comes from the secondary star, it must come from its irradiated face, which is closer to the centre of mass of the binary so will have a velocity amplitude smaller than $K_2$. The most extreme possibility is where the emission comes from Roche L1 point, in which case the true $K_2$ is related to the observed $K_2^{\rm\,obs}$ by
\begin{equation} K_2 = \frac{K_2^{\rm\,obs}}{(1+q)r_{\rm L1} - q} \end{equation}
where $r_{\rm L1}$ is the radius of the inner Lagrangian point in units of the semimajor axis. For $q = 0.15$ and $K_2^{\rm\,obs} = 202$\kms, this leads to $K_2 = 317$\kms. This in turn gives $r_{\rm AD} = 0.29 > r_{\rm h}$, allowing the implied disc radius to exceed the circularisation radius.

\begin{figure} \centering \includegraphics[height=0.48\textwidth,angle=270]{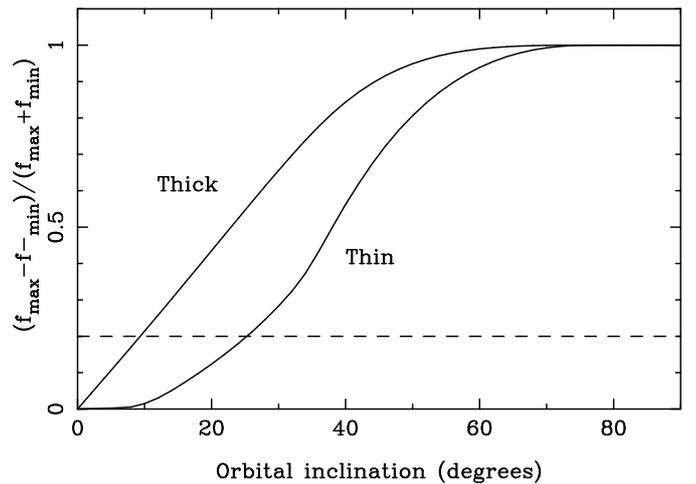} \\
\caption{\label{fig:modplot} The modulation of irradiation-induced line emission
from the secondary star versus orbital inclination. Both optically thin and optically
thick cases are indicated. The dashed line indicates the level of modulation of the
central sinusoidal component seen in the data.} \end{figure}

But there is another problem. If the central spike emission is coming from the irradiated face of the donor, it should peak in strength at orbital phases 0.25--0.75 when the face is pointing towards the observer (phase zero is defined by superior conjunction of the WD). The level of modulation will depend on inclination, with edge-on systems showing 100\% modulation and face-on systems no modulation. The brightness modulation measured by our four-Gaussian model (above) is $20 \pm 4$\% in amplitude. This quantity is subject to unknown systematics since the central spike is crossed by the bright spot and our fit is far from perfect, but it is clear that the modulation is not strong.

We then calculated the emission line to be expected from the irradiated face of the secondary by computing a grid covering the star and assigning an emissivity to each element proportional to the flux it receives from the WD and boundary layer. We calculated the modulation ($m$) from the minimum and maximum fluxes at orbital phases 0.0 and 0.5 ($f_\mathrm{min}$ and $f_\mathrm{max}$) using
\begin{equation} m = \frac{f_\mathrm{max} - f_\mathrm{min}}{f_\mathrm{max} + f_\mathrm{min}} \end{equation}
This was done both for optically thick emission and for the less likely possibility of optically thin emission. The results are plotted versus orbital inclination in Fig.~\ref{fig:modplot}. They show that if the central spike comes from the irradiated face of the secondary star, the orbital inclination must be low ($i < 10^\circ$ for optically thick and $i < 25^\circ$ for optically thin emission).

\begin{figure} \includegraphics[width=0.48\textwidth,angle=0]{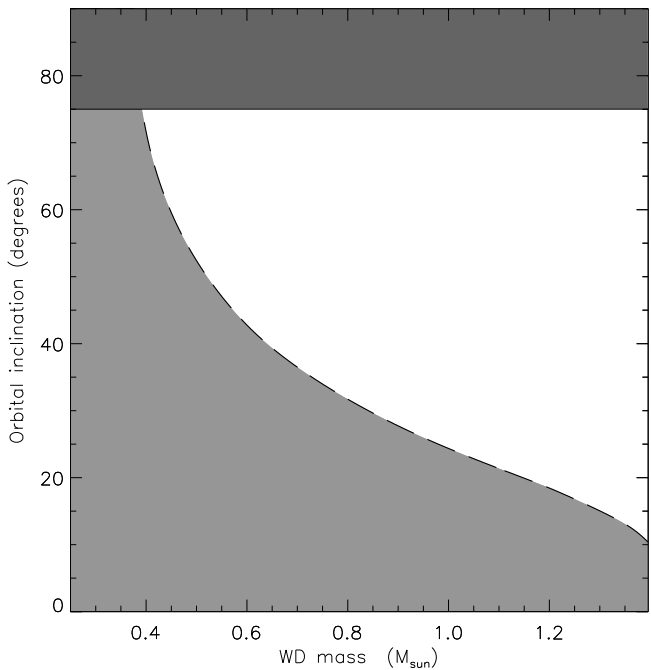} \\
\caption{\label{fig:iconstraint} The possible orbital inclinations of SDSS\,J0039
as a function of WD mass. The upper shaded region is discounted as the system does
not eclipse. The lower shaded region can be rejected using the measured FWZI of the
H$\alpha$ emission line.} \end{figure}

We can obtain a second constraint on the orbital inclination from the full width at zero intensity (FWZI) of the H$\alpha$ emission line profile, which we measure to be $2000 \pm 100$\kms. With a mass-radius relation for WDs [\citet{Bergeron++95apj}, supplemented by \citet{HamadaSalpeter61apj} for masses 1.2--1.4\Msun], the FWZI gives an upper limit on $i$ (as a function of WD mass) for which the innermost part of the disc is on a Keplerian orbit which is not inside the WD surface. We can also rule out high inclinations ($i \ga 75^\circ$; \citealt{Hellier01book}) as the system is not eclipsing (Sect.\,\ref{sec:phot}). These two constraints are visualised in Fig.\,\ref{fig:iconstraint} and show that requiring $i < 10^\circ$ ($25^\circ$) leads to WD masses of $M_{\rm WD} \ga 1.40$\Msun\ ($0.97$\Msun).


\subsection{True versus observed $K_2$ values}

If the central spike comes from optically thin gas situated at the Roche L1 point, then $i < 25\degr$ and $M_{\rm WD} \ga 0.97$\Msun. The validity of these constraints can be tested using simulated spectra of the system in order to connect $K_{\rm spike}$ with the true $K_2$. We therefore computed synthetic H$\alpha$ emission lines, for specific $q$ and $K_{\rm WD}+K_2$ values, as would arise from optically thin emission from the L1 point. These were measured for velocity and fitted with a sinusoid to obtain $K_2^{\rm\,obs}$. For ($q$,$K_{\rm WD}+K_2$) $=$ ($0.15$,$307$) we obtain $K_2^{\rm\,obs} = 202$\kms, and $K_2^{\rm\,true} = 261$\kms. This leads to $M_{\rm WD} \ga 2.2$\Msun\ so must be rejected on the grounds that the WD mass should not be above the Chandrasekhar limit. We can bring $M_{\rm WD}$ below the Chandrasekhar limit by adopting $q \la 0.06$, but such mass ratios result in an accretion disc which is far too small (see Fig.\,\ref{fig:qconstraints}).

It is marginally possible to satisfy all the above constraints by pushing parameter values to their limits: adopting $i < 30^\circ$ and $q = 0.15$ yields $M_{\rm WD} \la 1.33$\Msun\ and $r_{\rm AD} = 0.21$ greater than $r_{\rm h} = 0.19$. However, Doppler maps of this contrived configuration fail to correctly reproduce the position of the bright spot in velocity space.

To summarise this section, we find that the central spike cannot represent the centre of mass of the secondary star because this would result in an accretion disc which is too small. It cannot come from the L1 point: this either still causes the acccretion disc to be too small or the WD to be above Chandrasekhar mass. The best-compromise set of parameters are still very unlikely and also fail to explain the Doppler map of the system. We are therefore unable to attribute the central H$\alpha$ emission spike to either the accretion disc, the WD, or the secondary star.

\section{A slingshot prominence?}

We have discounted the WD and the secondary star as the source of the central spike. The accretion disc can be disregarded because it is incapable of producing such a narrow emission line. We can also rule out the bright spot where the disc encounters the mass transfer stream from the secondary star, which has a much higher velocity amplitude of roughly 1000\kms. There is one remaining possibility which cannot be rejected: a slingshot prominence (coronal loop) from the magnetically active secondary star. The existence of such a phenomenon was suggested from the observations of stationary central components in the spectra of IP\,Peg and SS\,Cyg in outburst \citep{Steeghs+96mn}. Some evidence for these prominences has also been presented for the long-period CV BV\,Cen \citep{Watson+07mn} and the magnetic system AM\,Her \citep{Gansicke+98aa,Kafka+08apj}.

There are, however, several differences between these objects and SDSS\,J0039. Firstly, the velocity amplitude of the slingshot prominence for SDSS\,J0039 is $202 \pm 3$\kms, whereas for the other CVs the emitting material is almost stationary. This motion is clear evidence that the emitted flux comes from within the SDSS\,J0039 system, in agreement with observations of IP\,Peg which show that the prominence is periodically eclipsed by the secondary star.

Secondly, the slingshot prominence in SDSS\,J0039 was observed during quiescence, when the system was in a state of low mass transfer. This contrasts with IP\,Peg and SS\,Cyg, where the prominences were seen during outburst, and with BV\,Cen, a long-period system which maintains a high mass transfer rate. Slingshot prominences originate from magnetic activity in the secondary star so are not directly related to the mass transfer rate, but when the rate is high there will be a greater photon flux and so more chance of seeing emission from the material in the prominence. The case of AM\,Her is different again, as it contains a strongly magnetic WD.

Thirdly, the longer orbital periods of SS\,Cyg (396\,min), IP\,Peg (228\,min), and BV\,Cen (880\,min) compared to SDSS\,J0039 (91.4\,min) imply very different secondary star properties. Whereas longer-period CVs contain more massive stars which are expected to be magnetically very active, shorter-period CVs possess fully convective secondaries with fundamentally different magnetic field configurations \citep{VerbuntZwaan81aa,Donati+06sci,Hallinan+06apj}. These secondary stars will also be rotating extremely quickly, as tidal effects force the rotation period to match the orbital period.


\section{Summary}

We have presented time-resolved VLT/FORS2 spectroscopy of the cataclysmic variable SDSS\,J003941.06$+$005427.5, which shows a strong triple-peaked H$\alpha$ emission line. From measurements of the H$\alpha$ line wings we found an orbital period of $91.395 \pm 0.093$\,min, making SDSS\,J0039 one of the mounting number of faint short-period CVs which have been identified by the SDSS \citep{Me+06mn,Gansicke+09mn}. From analysis of its spectral energy distribution we constrained the spectral type of the secondary star to be later than approximately L2. The semi-empirical donor sequence constructed by \citet{Knigge06mn} gives the spectral type of the mass donor of a 91\,min CV to be `M6.6' before and `T' after reaching the minimum orbital period \citep[see][]{Littlefair+08mn}, making SDSS\,J0039 a promising candidate to be a post-bounce CV.

We have decomposed the structures of the H$\alpha$ and He\,I 6678\,\AA\ and 7065\,\AA\ lines into velocity space using the Doppler tomography technique. The resulting Doppler maps show several oddities: an apparently elliptical accretion disc and a strong central emission feature with a velocity amplitude of $202 \pm 3$\kms. Our dynamical arguments show that this `central spike' has a velocity which is too high to be ascribed to the white dwarf and too low to arise from the surface of the secondary star. In addition to this, it shows a flux modulation of $20 \pm 4$\% on the orbital period of SDSS\,J0039, which also rules out the slim possibility that the emission comes from the inner Lagrangian point of the Roche-lobe-filling secondary star.

We can also discount the accretion disc and its bright spot as the source of the central spike, leaving us with one remaining possibility: a slingshot prominence rooted in the magnetically active secondary star. If confirmed, SDSS\,J0039 would represents the first observation of possible large-scale magnetic activity in a CV with an orbital period below the 2--3\,hr gap apparent in the period distribution of these objects. This means that the source star is of very low mass, fully convective \citep{MullanMacdonald01apj}, and rapidly rotating due to tidal effects. An addition difference is that the slingshot prominence has a substantial velocity amplitude compared to the almost-stationary prominences seen in longer-period CVs.

Slingshot prominences are expected to be short-lived phenomena, so this hypothesis could be tested by spectroscopically monitoring SDSS\,J0039 over several years.


\begin{acknowledgements}

Based on observations made with ESO Telescopes at the La Silla or Paranal Observatories under programme ID 079.D-0024. The reduced spectra, light curves and radial velocity observations presented in this work will be made available at the CDS ({\tt http://cdsweb.u-strasbg.fr/}) and at {\tt http://www.astro.keele.ac.uk/$\sim$jkt/}. JS, TRM, BTG and CMC acknowledge financial support from STFC in the form of grant number ST/F002599/1. DS acknowledges an STFC Advanced Fellowship. The following internet-based resources were used in research for this paper: the ESO Digitized Sky Survey; the NASA Astrophysics Data System; the SIMBAD database operated at CDS, Strasbourg, France; and the ar$\chi$iv scientific paper preprint service operated by Cornell University.


\end{acknowledgements}


\bibliographystyle{aa}

\end{document}